\documentclass[aps,twocolumn,showpacs]{revtex4}
\usepackage{amssymb,amsmath}

\begin{document}

\title{High-energy effective theory for orbifold branes}

\author{Tetsuya Shiromizu$^{(1,2,3)}$, Shunsuke Fujii$^{(1)}$, Claudia de Rham$^{(4)}$ and Hirotaka 
Yoshino$^{(1)}$}


\affiliation{$^{(1)}$Department of Physics, Tokyo Institute of Technology, Tokyo 152-8551, Japan}

\affiliation{$^{(2)}$Department of Physics, The University of Tokyo,  Tokyo 113-0033, Japan}

\affiliation{$^{(3)}$Advanced Research Institute for Science and Engineering,
Waseda University, Tokyo 169-8555, Japan}

\affiliation{$^{(4)}$Ernest Rutherford Physics Building, McGill
  University, 3600 rue University, Montreal, QC H3A 2T8, Canada}

\date{\today}


\begin{abstract}
We derive an effective theory on the orbifold branes of the
Randall-Sundrum 1 (RS1) braneworld scenario in the presence of a bulk brane. We 
concentrate on the regime
where the three branes are close and consider a scenario where the
bulk brane collides with one of the orbifold branes. This
theory allows us to understand the corrections to a low-energy
approach due to the presence of higher velocity terms, coming from
the Kaluza-Klein modes. We consider the evolution of gravitational 
waves on a cosmological background
and find that, within the large velocity limit, the boundary branes
recover a purely four-dimensional behavior. 
\end{abstract}

\maketitle



Recently, new scenarios have been suggested for a potential explanation for the 
creation of a hot big bang Universe 
\cite{CU}. The realization of such scenarios strongly depends on 
the possibility of a brane collision embedded in a
higher-dimensional bulk. Motivated by this idea, an effective theory
for a close brane system was addressed in
Refs. \cite{CL1,CL2,CL3,DFSY}(See Refs. \cite{collision, Bubble} for related issues).
The procedure used in these papers has been developed in order to 
consider the high energy corrections to the low-energy theory 
within a close-brane regime. In this regime,
we assume that the separations between the branes are much shorter than
the bulk curvature scale. Recently cosmological perturbations was also
discussed in
\cite{MT}, where an expansion in the brane velocities has instead
been proposed.
As a specific example,
the effective theory on a bulk brane colliding with the orbifold branes has been
studied in our previous paper \cite{DFSY}. The bulk brane 
has numerous specific features due to the reflection asymmetry across the bulk brane. 
In the ekpyrotic scenario \cite{CU}, however,
our Universe is assumed to be on one of the orbifold branes. Thus 
the aim of this report is to extend our previous study to the 
discussion for the boundary branes when such a bulk brane is.

For this purpose, we first describe the system in detail. Then we 
derive the effective theories on both orbifold branes and consider 
the evolution of tensor perturbations. We show 
that the tensor perturbations follow an equation which is
different from what would be expected from a four-dimensional
low-energy approach.

We consider the RS1 model \cite{RSI} 
with an additional bulk brane between the two orbifold branes. 
The reflection symmetry is imposed for the two orbifold branes. 
On the other hand, no such reflection symmetry is 
imposed across the bulk brane \footnote{The precise description of our setup 
is seen from the junction condition for the extrinsic curvature described 
below.} (See Refs. \cite{Davis:2000jq} for general
issues on reflection asymmetric branes). 
The five dimensional metric of the bulk spacetime on both side of the bulk brane is
%
\begin{eqnarray}
ds^2_\pm=e^{2\varphi_\pm (y,x)}dy^2+g_{\mu\nu}(y,x)dx^\mu dx^\nu,\label{ds2}
\end{eqnarray}
%
where $y$ is the coordinate of extra dimension, and the index $+$
(resp. $-$)
represents the bulk region located between the bulk brane and the
positive (reps. negative) tension orbifold brane, which is located at
$y=y_+$ (resp. $y=y_-$).
The location of the bulk brane is $y=y_0$ and we can suppose 
$y_+ < y_0 < y_-$ without loss of generality. The gauge in \eqref{ds2} is
chosen such that the branes are fixed to these given locations.
The system is governed by the Einstein-Hilbert
action with negative cosmological constant $\Lambda_\pm=- 6/\kappa^2
  \ell_\pm^2$, $\ell_\pm$ being the bulk Anti-de Sitter  curvature.
The brane action is the Nambu-Goto one with the 
four-dimensional action for matter fields localized on that brane.
Due to the presence of boundary, the Gibbons-Hawking terms should also be introduced \cite{GH}.


In the frame \eqref{ds2}, 
the modified Einstein equation on the orbifold branes is written as 
\cite{CL1,SMS}
%
\begin{eqnarray}
{}^{(4)}R^\mu_\nu (y_I) & = &
-\frac{4}{\ell_I^2}\delta^\mu_\nu +K(y_I)K^\mu_\nu(y_I) +
e^{-\varphi_I} \partial_y K^\mu_\nu (y_I) \nonumber \\
& & +e^{-\varphi_I} D^\mu D_\nu e^{\varphi_I}|_{y_I}, \label{Gauss Coddaci}
\end{eqnarray}
%
where $K^\mu_\nu$ is the extrinsic curvature defined by $K^\mu_\nu = \frac{1}{2}
e^{-\varphi_I}g^{\mu\alpha}\partial_y g_{\alpha\nu}$ 
and the index $I=\pm$ 
represents the positive or negative 
tension brane.
$D_\mu$ is the covariant derivative with respect to the induced
metric $g_{\mu\nu}(y_I,x)$ on the
brane.
For the orbifold branes, the presence of the reflection symmetry allows us
to specify uniquely the extrinsic curvature $K^\mu_\nu (y_I)$ through the Isra\"el matching conditions
\cite{Israel:1966rt}. These junction conditions relate the
extrinsic curvature to the stress energy tensor ${}^{(I)}T^\mu_\nu$ of matter fields localized on the
brane:
%
\begin{eqnarray}
K^\mu _\nu(y_I) = - \frac{1}{\ell_I} \delta^\mu_\nu
-I \frac{\kappa^2}{2} \Bigl({}^{(I)}T^\mu_\nu -\frac{1}{3}\delta^\mu_\nu
{}^{(I)}T \Bigr).
\end{eqnarray}
%
In what follows, we will focus on the evaluation of $\partial_y K^\mu_\nu (y_I)$ in \eqref{Gauss
Coddaci},
which is the only unknown quantity. Since we are interested in a
close-brane limit, we will use a Taylor expansion for the current system. 
We first note that, using Taylor expansion around
$y=y_I$ and the recursive relation 
$\partial_y^n K^\mu_\nu(y_I)=\hat O_I 
\partial_y^{n-2}K^\mu_\nu (y_I) (y_I-y_0)^{-2}$, 
$K^\mu_\nu (y_0)$ can be written as \cite{CL2,CL3,DFSY}
%
\begin{eqnarray}
{}^{I}K^\mu_\nu (y_0) & = & {\rm cosh}{\sqrt {\hat O_I}}K^\mu_\nu (y_I)
\nonumber \\
& & +(y_0-y_I)\frac{{\rm sinh}{\sqrt {\hat O_I}}}{{\sqrt {\hat O_I}}}
\partial_y K^\mu_\nu (y_I),
\end{eqnarray}
%
where the operator $\hat O_I$ is defined below. 
This gives us the following expression for the derivative of the
extrinsic curvature:
%
\begin{eqnarray}
e^{-\varphi_I} \partial_y K^\mu_\nu (y_I) & = & \frac{I}{d_I}
\frac{{\sqrt {\hat O_I}}}{{\rm sinh}{\sqrt {\hat O_I}}}
\Biggl({}^I K^\mu_\nu(y_0) \nonumber \\
& & -{\rm cosh}{\sqrt {\hat O_I}}K^\mu_\nu(y_I) \Biggr), \label{K'}
\end{eqnarray}
%
where $d_I$ represents the distance between the bulk brane and the
orbifold brane:
%
\begin{eqnarray}
d_I \equiv e^{\varphi_I (x)} \  |y_0-y_I|
\end{eqnarray}
%
and ${}^{\pm}K^\mu_\nu (y_0):=\lim_{\epsilon \to 0}
K^\mu_\nu (y_0 \mp \epsilon)$.
The action of the operators $\hat{O}$ on any four-dimensional symmetric tensors $S_{\mu
\nu}$ are defined by 
%
\begin{eqnarray}
\hat{O}_{IJ} & \equiv & |(y_0-y_I)(y_0-y_J)| \hat{U}_{IJ} \qquad I,J=\pm \label{O IJ}\\
\hat{U}_{IJ} S^\mu _\nu &=& D^\mu e^{\varphi_I} D_\alpha e^{\varphi_J} S^\alpha_\nu
+D_\nu e^{\varphi_I} D_\alpha e^{\varphi_J}   S^{\mu \alpha } \nonumber \\
& & -D_\alpha e^{\varphi_I} D^\alpha e^{\varphi_J} S^\mu_\nu. 
\end{eqnarray}
%
For the brevity, we used the notation of $\hat O_{II}=\hat O_I$. In general, 
$|D_\mu d_I|$ is of order $|D_\mu d_I| = \mathcal{O}(d_I^0)$, (See
Refs. \cite{CL2,CL3} for details). The operator $\hat O_{IJ}$ is
therefore of the same order:
$\hat O_{IJ}= \mathcal{O}(Dd_I\, Dd_J) = \mathcal{O}(d^0_I d^0_J)$. Note that the operator 
$\hat O_{IJ}$ here 
is defined with respect to the orbifold brane metric.

In the expression \eqref{K'}, ${}^IK^\mu_\nu(y_0)$ is still unknown 
quantity. Let then decompose ${}^{I}K^\mu_\nu (y_0)$ into two parts as  
%
\begin{eqnarray}
{}^{I}K^\mu_\nu (y_0) = -I \frac{1}{2}\Delta K^\mu_\nu (y_0)
+\bar
K^\mu_\nu (y_0), \label{Kbulk}
\end{eqnarray}
%
where $\Delta K^\mu_\nu(y_0) $ is the part which can be determined by the
junction condition at $y=y_0$:
%
\begin{eqnarray}
\Delta K^\mu_\nu(y_0) & := & {}^{-}K^\mu_\nu(y_0)-{}^{+}K^\mu_\nu(y_0) \nonumber \\
& = & \left(\frac{1}{\ell_+}-\frac{1}{\ell_-}\right)\delta^\mu_\nu
- \kappa^2 \Bigl({}^{(0)}T^\mu_\nu -\frac{1}{3}\delta^\mu_\nu {}^{(0)}T
\Bigr). \label{Bulk Junction}
\end{eqnarray}
%
In the above ${}^{(0)}T^\mu_\nu$ is the energy-momentum tensor of matter fields localized on the
bulk brane. $\bar K^\mu_\nu (y_0)$ is the averaged part  (or ``asymmetric" term) defined by 
%
\begin{eqnarray}
\bar K^\mu_\nu (y_0):=\frac{1}{2}\Bigl({}^{+}K^\mu_\nu(y_0)+{}^{-}K^\mu_\nu
(y_0)  \Bigr).
\end{eqnarray}
%
Using the result of Ref. \cite{DFSY},  in the
close-brane regime, the asymmetric term can be expressed as:
%
\begin{eqnarray}
\bar K^\mu_\nu (y_0)= \hat L \ Z^\mu_\nu, \label{K bar}
\end{eqnarray}
%
where
%
\begin{eqnarray}
\hat L = -\left[\frac{1}{d_-}\frac{{\sqrt {\hat O_-}}}{{\rm tanh}{\sqrt {\hat O_-}}}
+\frac{1}{d_+}\frac{{\sqrt {\hat O_+}}}{{\rm tanh}{\sqrt {\hat O_+}}} \right]^{-1}
\label{L operator}
\end{eqnarray}
%
and
%
\begin{multline}
Z^\mu_\nu  \equiv \sum_{I=\pm} \Biggl[ \frac{I}{d_I}D^\mu D_\nu |_{y_0}
d_I-\frac{1}{d_I}\frac{\sqrt{\hat{O}_I}}{\sinh \sqrt{\hat{O}_I}}\
K^\mu_\nu(y_I) \label{average}\\
-\frac{I}{2 d_I}
\frac{\sqrt{\hat{O}_I}}{\tanh \sqrt{\hat{O}_I}}\
\Delta K^\mu_\nu(y_0)\Biggr]. 
\end{multline}
%
In Eqs. \eqref{L operator} and \eqref{average}, the operator $\hat
O_I$ is defined on the bulk brane as $\hat O_I := (y_0-y_I)^2
\, \hat U_{II}$, 
where $D_\mu$ is the covariant derivative
with respect to the bulk brane metric. Since we are working in a close-brane regime,
$g_{\mu \nu}(y_0) = g_{\mu \nu}(y_\pm) +\mathcal{O}(d)$, and then
%
\begin{eqnarray}
\hat{O} |_{y_0} =  \hat{O} |_{y_I} + \mathcal{O}(d). \label{OO}
\end{eqnarray}
%
In the current approximation, we can therefore simply replace $\hat O$ in the expression for $\bar
K^\mu_\nu (y_0)$ by
the operator on the orbifold branes.

The remaining task to obtain the effective theory on the orbifold branes is to 
write down $D^\mu D_\nu d_I|_{y_0} $ in terms of quantities on
the orbifold branes. This can be performed using the Taylor expansion:
%
\begin{equation}
D^\mu D_\nu e^{\varphi_I}|_{y=y_0}=\sum_{n\ge 0}
\frac{\left(y_J-y_0\right)^n}{n!}\ \partial_y^n (D^\mu D_\nu e^{\varphi_I})|_{y=y_J}, \label{taylor
Ddd}
\end{equation}
%
with
%
\begin{equation}
\partial _y^{2n}(D^\mu D_\nu e^{\varphi_I})|_{y=y_J}=
-\hat U_{JI} \hat U_{J}^{n-1} \partial_y K^\mu_\nu (y_J) \label{Ric even}
\end{equation}
%
and
%
\begin{equation}
\partial_y^{2n-1}(D^\mu D_\nu e^{\varphi_I})|_{y=y_J}=
- \hat U_{JI} \hat U_{J}^{n-1} K^\mu_\nu (y_J). \label{Ric odd}
\end{equation}
%
Eqs. \eqref{Ric even} and \eqref{Ric odd} are derived from the
expression for derivative of the Christoffel symbol(See Ref. \cite{CL2} 
for the detail). As a result, one has
\\
%
\begin{eqnarray}
&& \frac{1}{d_I}D^\mu D_\nu d_I |_{y=y_0}=
\frac{1}{d_I}D^\mu D_\nu d_I |_{y=y_J} \nonumber \\
& &\hspace{20pt}
-\frac{1}{d_I}\hat O_{JI} \frac{{\rm cosh}{\sqrt {\hat O_J}}-1}{\hat O_J}
\Bigl( d_J R^\mu_\nu (y_J)-D^\mu D_\nu d_J|_{y_J}  \Bigr)\nonumber \\
&&\hspace{20pt}-J \frac{1}{d_I} \hat O_{JI} \frac{{\rm sinh}{\sqrt
{\hat
O_J}}}{{\sqrt {\hat O_J}}}
K^\mu_\nu (y_J),  \label{ij}
\end{eqnarray}
%
where we used the equation 
%
\begin{eqnarray}
\partial _y K^\mu {}_\nu = e^\varphi  {}^{(4)}R^\mu{}_\nu -D^\mu D_\nu e^\varphi +\mathcal{O}(d). 
\label{del y
K}
\end{eqnarray}
%
Substituting Eq. \eqref{ij} into Eq. \eqref{K bar}(with Eq. \eqref{Kbulk}), 
we can write down the extrinsic curvature on the bulk brane in terms of 
the quantities on the orbifold branes. Then we can compute the derivative 
of the extrinsic curvature on the orbifold branes, $\partial_y K^\mu_\nu (y_I)$, 
from Eq. (\ref{K'}). All ingredients for the derivation 
of the close-brane effective theory \eqref{Gauss Coddaci} are therefore presented. 

But first, one may concentrate on the trace part of the four-dimensional Ricci tensor 
which is determined through the Hamiltonian constraint as 
%
\begin{equation}
{}^{(4)}R(y_I) = -K^{\alpha}_\beta (y_I) K_\alpha^\beta (y_I)
+K^2(y_I)-\frac{12}{\ell^2_I} =\mathcal{O}(d^0). 
\end{equation}
%
Then it is turned out to be 
negligible compared to the traceless part of the Ricci tensor 
because the traceless part is order of $\mathcal{O}(d^{-1})$. 

The equation for the radion may be derived using the tracelessness property
of the projected Weyl tensor $E_{\mu\nu}(y_I):={}^{(5)}C^y_{~~\mu y \nu}(y_I)$. 
On the orbifold branes, $E^\mu_\nu(y_I)$ becomes \cite{SMS} 
%
\begin{eqnarray}
E^\mu_\nu(y_I)& = & -e^{-\varphi_I}\partial_y K^\mu_\nu (y_I)
-e^{-\varphi_I} D^\mu D_\nu e^{\varphi_I} |_{y_I} \nonumber \\
& & +\mathcal{O}(d^0).
\end{eqnarray}
%
Using the fact of $E^\mu_\mu(y_I)=0$, we obtain the Klein-Gordon equation for the radion 
$d_I$ as 
%
\begin{eqnarray}
D^2d_I|_{y_I} & = & -I \delta^\nu_\mu
\frac{{\sqrt {\hat O_I}}}{{\rm sinh}{\sqrt {\hat O_I}}}
\Biggl({}^{I} K^\mu_\nu (y_0) \nonumber \\
& & -{\rm cosh}{\sqrt {\hat O_I}}K^\mu_\nu (y_I) \Biggr).
\end{eqnarray}
%
Finally the effective theory on the orbifold branes is derived as
%
\begin{eqnarray}
{}^{(4)}G^\mu_\nu (y_I) & = & \frac{1}{d_I}
\left(\delta^\mu_\beta \delta^\alpha_\nu-\delta^\mu_\nu \delta^\alpha_\beta\right)
\Biggl[
 D^\beta D_\alpha d_I |_{y_I} \nonumber \\
& &\hspace{-60pt}
+ I \frac{{\sqrt {\hat O_I}}}{{\rm \sinh}{\sqrt {\hat O_I}}} 
\Biggl({}^{I} K^\beta_\alpha (y_0)
-{\rm cosh}{\sqrt {\hat O_I}}K^\beta_\alpha (y_I)
\Biggr)
\Biggr], \label{eff theory}
\end{eqnarray}
%
where ${}^{I} K^\mu_\nu (y_0)$ can be expressed in terms of
$Z^\mu_\nu$ and $\Delta K^\mu_\nu (y_0)$ using Eqs. (\ref{Kbulk}) and (\ref{K bar}). 
The tensor $Z^\mu_\nu$  can be written in the terms of $\Delta
K^\mu_\nu (y_0)$ and $K^\mu_\nu (y_I)$. For that we
substitute Eq. \eqref{ij} for $\frac{1}{d_I}D^\mu D_\nu d_I
|_{y=y_0}$ into the expression \eqref{average} for $Z^\mu_\nu$ and then 
%
\begin{eqnarray}
\hspace{-30pt}Z^\mu_\nu &\hspace{-2pt} = &\hspace{-2pt} \sum_{J=\pm} \frac{1}{d_j}\Biggl[J
\Biggl\lbrace D^\mu D_\nu d_J |_{y_I}
- I  \hat O_{IJ} \frac{{\rm sinh}{\sqrt {\hat O_I}}}{{\sqrt {\hat
O_I}}}
K^\mu_\nu (y_I)\nonumber \\
&\hspace{-2pt}- &\hspace{-2pt} \hat O_{IJ} \frac{{\rm cosh}{\sqrt
{\hat
O_I}}-1}{\hat
O_I}
\Bigl( d_I R^\mu_\nu (y_I)-D^\mu D_\nu d_I |_{y_I} \Bigr)\Biggr\rbrace\nonumber \\
&\hspace{-2pt} -&\hspace{-2pt}
\frac{\sqrt{\hat{O}_J}}{\sinh
\sqrt{\hat{O}_J}}
K^\mu_\nu(y_J) 
- J
\frac{\sqrt{\hat{O}_J}}{2\tanh \sqrt{\hat{O}_J}}
\Delta K^\mu_\nu(y_0)\Biggr].
\end{eqnarray}
%
The effective theory \eqref{eff theory} is the main result of this 
brief report and may be used for any studies within the close-brane 
regime. In particular this regime is of interest if one considers a bulk brane 
colliding with a boundary brane, and wishes to understand the propagation of the 
perturbations throughout this process. 
As we shall see below the behavior is different from the
low-energy prescription.


As a concrete example, we consider the branes to be empty at the
background 
level (except for a canonical tension) such that ${}^{(\hat I)} T^\mu_\nu=0$, for $\hat I=\pm,0$. We 
then study the 
evolution of tensor perturbations generated by small stress-energy tensor  ${}^{(\hat I)}\delta T^i_j$ on the branes at the perturbed
level. The sources ${}^{(\hat I)}\delta T^i_j$ are taken to be
tensor-like such that ${}^{(\hat I)}\delta T^j_j=0$ and ${}^{(\hat I)}\delta
T^i_{j,i}=0$.
We take the metric on the orbifold branes to be
%
\begin{eqnarray}
ds^2 & = & -dt_I^2+a_I^2(t)\Bigl(\delta_{ij}+ h_{ij}(y_I)\Bigr)dx^i dx^j \nonumber \\
     & = & a_I^2(\tau)\Bigl[-d \tau_I^2+\Bigl(\delta_{ij}+h_{ij}(y_I)\Bigr)dx^i dx^j \Bigr],
\end{eqnarray}
%
where $t_I$ and $\tau_I$ are the cosmic time and conformal time on
the orbifold branes. We may point out that $\hat O_+$ and $\hat O_-$ commute
for linear perturbations as pointed out in \cite{DFSY}.
The linear part of Ricci tensor is
%
\begin{eqnarray}
\delta R^i_j(y_I) & = &  \frac{1}{d_I} \delta (D^i D_j d_I )
+I \frac{{\sqrt {O}_I}}{{\rm sinh}{\sqrt {\hat O}_I}}
\Bigl(\delta {}^{I}K^i_j (y_0) \nonumber \\
& & -{\rm cosh}{\sqrt {\hat O}_I} \delta K^i_j(y_I) \Bigr),
\end{eqnarray}
%
which leads us to the linearized equation for the
gravitational waves $h^i_j(y_\pm)$ on the orbifold branes after long but straight-forward 
calculations:
%
\begin{eqnarray}
\hat \boxdot_\pm h^i_j(y_\pm) = -\kappa^2 a_\pm^2 \Omega_\pm  {}^{({\rm eff},\pm)}\delta T^i_j,
\end{eqnarray}
%
where
%
\begin{eqnarray}
& \hat \boxdot_\pm := \eta^{\mu\nu}\partial_\mu \partial_\nu
-\frac{1}{U_\pm}\frac{d_+'d_-'}{d_+d_-}\Bigl(\frac{1}{{\rm tanh}\dot d_+}+\frac{1}{{\rm tanh}\dot d_-}
\Bigr)\partial_{\tau_\pm}\\
& U_\pm=\frac{\dot d_\pm}{d_\pm} \Bigl(\frac{1}{{\rm tanh}\dot d_+}+\frac{1}{{\rm tanh}\dot d_-} 
\Bigr)
 \pm \frac{1}{{\rm sinh}\dot d_\pm} \Bigl(\frac{\dot d_+}{d_+}-\frac{\dot d_-}{d_-} \Bigr)\\
& \Omega_\pm = \frac{1}{U_\pm} \frac{\dot d_+}{d_+} \frac{\dot d_-}{d_-},
\end{eqnarray}
%
and the effective stress-energy ${}^{({\rm eff},\pm)}\delta T^i_j$ is
given in terms of the stress-energy tensors
${}^{(\hat I)}\delta T^i_j$ on the branes:
%
\begin{eqnarray}
{}^{({\rm eff},\pm)}\delta T^i_j & = &
\Biggl(1+ \frac{1}{{\rm tanh}\dot d_+ {\rm tanh} \dot d_-}  \Biggr) {}^{(\pm)}\delta T^i_j \nonumber
\\
& & +\frac{1}{{\rm sinh}\dot d_+ {\rm sinh} \dot d_-}
{}^{(\mp)}\delta T^i_j \nonumber \\
& & + 2\frac{{\rm cosh}\dot d_\mp}{{\rm sinh}\dot d_+ {\rm sinh} \dot d_- } {}^{(0)}\delta T^i_j.
\end{eqnarray}
%

In particular we may be interested in the large velocity limit ($|\dot d_I| \gg 1$).
In this limit, the linearized equation becomes
%
\begin{eqnarray}
\Biggl(\eta^{\mu\nu} \partial_\mu \partial_\nu -\frac{d'_\mp}{d_\mp}\partial_{\tau_\pm}
\Biggr) h^i_j(y_\pm)= - a_\pm^2 \kappa^2 \frac{\dot d_\mp}{d_\mp}
{}^{(\pm)}\delta T^i_j. \label{fast}
\end{eqnarray}
%
The perturbations on a given orbifold brane only depend on its
own matter content and do not depend on the ones on the bulk brane and
the other orbifold brane. Here note that the matter content of the other branes
contribute at small velocities. This is a remarkable 
result since this high-energy theory reproduces an exact four-dimensional
behavior without any backreaction from the presence of the other 
branes, which would not be the case in  a low-energy regime. 
In Eq. \eqref{fast}, one might be surprised by the presence of the
quantity $\dot d_\mp /d_\mp$ rather than $\dot d_\pm/d_\pm$. But
within the close-brane behavior, we may recall that $d_\pm =\dot d_\pm
\tau +\mathcal{O}(\tau ^2)$, 
and hence $\dot d_+ /d_+\sim \dot d_- /d_-\sim
1/\tau$, which is precisely the factor expected from a
four-dimensional theory \cite{CL3}.

As a consistency check, we may also consider the special case where 
the reflection symmetry is imposed
by hand across the bulk brane.
The two orbifold branes should be identified and one can set
$d_+=d_-=:d$, $a_+ = a_- = a$ and ${}^{(+)}\delta T^i_j ={}^{(-)}\delta
T^i_j=:\delta T^i_j $ so that the linearized equation for tensor
perturbations becomes
%
\begin{eqnarray}
&&\hspace{-30pt}\Biggl(\eta^{\mu\nu} \partial_\mu \partial_\nu
-\frac{d'}{d}\partial_{\tau_\pm}
\Biggr) h^i_j(y_\pm) \nonumber \\
&&\hspace{30pt}=  - \kappa^2 a^2 \frac{\dot d}{d}
\Biggl( \frac{1}{{\rm tanh}\dot d} \delta T^i_j
 +\frac{1}{{\rm sinh}\dot d} {}^{(0)}\delta T^i_j \Biggr),
\end{eqnarray}
%
which is consistent with the result obtained in Ref. \cite{CL3} which
explored a two-brane system with the reflection symmetry.

 Finally we summarize our current study. In a previous work we derived the
effective theory for a bulk brane in the context of the RS1 braneworld model.
In this brief report, we
derived the effective theory for orbifold branes within the same setup.
In the close-brane regime, the metric on the orbifold branes and the
bulk brane are trivially related to each other.
However, we would like to stress that the individual derivation of 
the effective theory for each brane is important for the physical 
interpretation on the time evolutions as well as for 
fixing the initial conditions. We have therefore derived 
separately the effective close-brane theory on the orbifold branes 
and have recovered the similar main features found for the bulk 
brane. The difference appears in the gravitational coupling. 
Nevertheless, within the large-velocity limit, the behavior of the branes 
becomes purely four-dimensional and decouples entirely from the other branes, 
which is not the case within a low-energy regime. 


\section*{Acknowledgements}

The work of TS was supported by Grant-in-Aid for Scientific
Research from Ministry of Education, Science, Sports and Culture of
Japan(No.13135208, No.14102004, No. 17740136 and No. 17340075) and
the Japan-U.K. Research Cooperative Program. CdR is funded by a grant
from the Swiss National Science Foundation.


\begin{thebibliography}{22}


\bibitem{CU}
J. Khoury, B. A. Ovrut, P. J. Steinhardt and N. Turok, Phys. Rev. {\bf D64}, 123522 (2001).


\bibitem{DFSY}
C. de Rham, S. Fujii, T. Shiromizu and H. Yoshino, Phys. Rev. {\bf D72}, 123522 (2005).


\bibitem{CL1}
T. Shiromizu, K. Koyama and K. Takahashi, Phys. Rev. {\bf D67}, 104011 (2003).

\bibitem{CL2}
C. de Rham and S. Webster, Phys. Rev. {\bf D71}, 124025 (2005).

\bibitem{CL3}
C.~de Rham and S.~Webster, Phys. Rev. {\bf D72}, 64013 (2005).

\bibitem{collision}
A. Neronov, JHEP {\bf 0111}, 007(2001);
D. Langlois, K. Maeda and D. Wands, Phys. Rev. Lett. {\bf 88},
181301 (2002);
S. Kanno, M. Sasaki and J. Soda, Prog. Theor. Phys. {\bf 109}, 357 (2003).


\bibitem{Bubble}
U. Gen, A. Ishibashi and T. Tanaka Phys. Rev. {\bf D66}, 023519 (2002);
J. J. Blanco-Pillado, M. Bucher, S. Ghassemi and F. Glanois, Phys. Rev. {\bf D69}, 103515 (2004).

\bibitem{MT}
P. L. MacFadden and N. Turok, hep-th/0512123.


\bibitem{RSI}
L.~Randall and R.~Sundrum, Phys. Rev. Lett. {\bf 83}, 3370 (1999).





\bibitem{Davis:2000jq}
  A.~C.~Davis, S.~C.~Davis, W.~B.~Perkins and I.~R.~Vernon,
  Phys. Lett. {\bf B504}, 254 (2001);
  %
  B.~Carter and J.~P.~Uzan,
  Nucl. Phys. {\bf B606}, 45 (2001);
%
  K.~Takahashi and T.~Shiromizu,
  Phys. Rev. {\bf D70}, 103507 (2004);
%
  L.~A.~Gergely, E.~Leeper and R.~Maartens,
  Phys. Rev. {\bf D70}, 104025 (2004);
%
  I.~R.~Vernon and D.~Jennings,
  JCAP {\bf 0507}, 011 (2005).


\bibitem{GH}
G. W. Gibbons and S. W. Hawking, Phys. Rev. {\bf D15}, 2752 (1977).


\bibitem{SMS}
T. Shiromizu, K. Maeda and M. Sasaki, Phys. Rev. {\bf D62},
024012 (2000),
%
  T.~Shiromizu and K.~Koyama,
  Phys.\ Rev.\ D {\bf 67}, 084022(2003)
  [arXiv:hep-th/0210066].



\bibitem{Israel:1966rt}
W.~Israel,
Nuovo Cim.\ B {\bf 44S10}, 1 (1966)
[Erratum-ibid.\ B {\bf 48} (1967\ NUCIA,B44,1.1966) 463].











\end{thebibliography}
\end{document}